\documentclass[aps,preprint,tightenlines,showpacs,nofootinbib]{revtex4}
\usepackage{epsfig}

\begin{document}

\title{Strangeness $-2$ two-baryon systems}
\author{A. Valcarce}
\affiliation{Departamento de F\'\i sica Fundamental, Universidad de Salamanca, E-37008
Salamanca, Spain}
\author{H. Garcilazo}
\affiliation{Escuela Superior de F\'\i sica y Matem\'aticas, 
Instituto Polit\'ecnico Nacional, 
Edificio 9, 07738 M\'exico D.F., Mexico}
\author{T. Fern\'andez-Caram\'es}
\affiliation{Departamento de F\'\i sica Fundamental, Universidad de Salamanca, E-37008
Salamanca, Spain}
\date{\today}

\begin{abstract}
We derive strangeness $-2$ baryon-baryon interactions from a 
chiral constituent quark model including the full set of scalar mesons. 
The model has been tuned in the strangeness $0$ and $-1$ two-baryon
systems, providing parameter free predictions for the strangeness $-2$
case. We calculate elastic and inelastic $N\Xi$ and $\Lambda\Lambda$ 
cross sections which are consistent with the existing experimental data.
We also calculate the two-body scattering lengths for the 
different spin-isospin channels.
\end{abstract}

\pacs{13.75.Ev,12.39.Jh,21.45.-v,21.80.+a}
\maketitle

\section{Introduction}
The knowledge of the strangeness $S=-2$ two-baryon interactions 
has become an important issue for theoretical and experimental 
studies of the strangeness nuclear physics. Moreover, this is
an important piece of a more fundamental problem, the description
of the interaction of the different members of the baryon octet
in a unified way. The $\Xi N -\Lambda \Lambda$ interaction accounts
for the existence of doubly strange hypernuclei, which is a
gateway to strange hadronic matter. Strangeness $-2$ baryon-baryon 
interactions also account for a possible six-quark $H$-dibaryon, 
which has yet to be experimentally observed.
Its knowledge is naturally invaluable for the quantitative
predictions for various aspects of neutron star matter.

There has been an steady progress towards the $S=-2$ baryon-baryon interaction.
The $\Lambda N$ interaction is pretty much understood based on the 
experimental data of $\Lambda$ hypernuclei. There is also 
some progress made on the $\Sigma N$ interaction. However,
the experimental knowledge on the $\Xi N$ and hyperon-hyperon ($YY$) interactions
is quite poor. The only information available came from doubly
strange hypernuclei, suggesting that the $^1S_0$ $\Lambda\Lambda$
interaction should be moderately attractive. An upper limit of
$B_{\Lambda \Lambda}=7.25\pm0.19$ MeV has been deduced for an 
hypothetical $H$-dibaryon (a lower limit of $2223.7$ MeV/$c^2$
for its mass) from
the so-called Nagara event~\cite{Nag08} at a 90\% confidence level.
The KEK-E176/E373 hybrid emulsion experiments observed other 
events corresponding to double$-\Lambda$ hypernuclei. Among
them, the Demachi-Yanagi~\cite{Nak09,Ich01}, identified as $^{10}_{\Lambda \Lambda}$Be
drove a value of $B_{\Lambda\Lambda}=11.90 \pm 0.13$~\cite{Hiy10}. A new
observed double-$\Lambda$ event has been recently reported by the
KEK-E373 experiment, the Hida event~\cite{Ich01},
although with uncertainties on its nature it has been
interpreted as an observation of the ground state 
of the $^{11}_{\Lambda \Lambda}$Be~\cite{Hiy10}.
Very recently, doubly strange baryon-baryon scattering data at
low energies were deduced for the first time, obtaining some
experimental data and upper limits for different elastic and inelastic 
cross sections: $\Xi^-p \to \Xi^- p$ and $\Xi^-p \to \Lambda \Lambda$~\cite{Ahn06}.
Recent results of the KEK-PS E522 experiment indicates the possibility
of a $H$-dibaryon as a resonance state
with a mass range between the $\Lambda\Lambda$ and
$N\Xi$ thresholds~\cite{Yoo07}. 

In the planned experiments at J-PARC~\cite{Nag00}, dozens of 
emulsions events for double$-\Lambda$ hypernuclei will be produced. 
The $(K^-,K^+)$ reaction is one of the most promising ways of studying
doubly strange systems. $\Lambda\Lambda$ hypernuclei can be produced
through the reaction $K^- p \to K^+ \Xi^-$ followed by $\Xi^- p \to \Lambda\Lambda$.
It is therefore compulsory having theoretical predictions concerning
the $\Xi N - \Lambda\Lambda$ coupling~\cite{Har10} to guide the
experimental way and, in a symbiotic manner, to use the new experimental data
as a feedback of our theoretical knowledge. Such a process may shed
light on our understanding of the nature of $SU(3)$ symmetry breaking
in the baryon-baryon interaction. Knowledge of the strangeness $-2$
sector is therefore a challenge for understanding
the baryon-baryon interaction in a unified way.
The chiral constituent quark model has been very successful in the
simultaneous description of the baryon-baryon interaction and the baryon 
spectrum as well as in the study of the two- and
three-baryon bound-state problem for the nonstrange sector~\cite{Val05}.
A generalization to the strange sector has been
applied to study the meson and baryon spectra~\cite{Gar05},
giving a nice description
of the hyperon-nucleon elastic and inelastic cross sections,
and valuable predictions for the strangeness
$-1$ three-baryon systems: $\Lambda NN -\Sigma NN$~\cite{GFV07}.

In this work, we will apply the same model to derive the strangeness
$-2$ baryon-baryon interactions: $\Lambda \Lambda$, $\Lambda\Sigma$, 
$\Sigma \Sigma$ and $N\Xi$. 
We will use these two-body interactions
to calculate two-body elastic
and inelastic scattering cross sections
and we will compare to experimental data and other theoretical models.
We will also calculate the two-body
scattering lengths of the different spin-isospin channels
to compare with other theoretical models. 
The structure of the paper is the following.
In the next section we will resume the basic aspects of 
the two-body interactions and we will present the
integral equations for the different two-body systems.
In section~\ref{Res} we present our results compared to
other models and the available experimental data.
Finally, in section~\ref{Sum} we summarize our main
conclusions.

\section{Formalism}
\label{Form}
\subsection{The strangeness $-2$ baryon-baryon potential}

The baryon-baryon interactions involved in the study of the coupled
$\Lambda\Lambda - N\Xi - \Lambda\Sigma - \Sigma\Sigma$  system are 
obtained from the chiral 
constituent quark model~\cite{Val05}. In this model baryons
are described as clusters of three interacting massive (constituent) quarks,
the mass coming from the spontaneous breaking of chiral symmetry. The
first ingredient of the quark-quark interaction is a confining
potential ($CON$). Perturbative aspects of QCD are taken into account
by means of a one-gluon potential ($OGE$). Spontaneous breaking of 
chiral symmetry gives rise to boson exchanges
between quarks. In particular, there appear pseudoscalar 
boson exchanges and their corresponding scalar partners~\cite{GFV07}. Thus,
the quark-quark interaction will read: 
\begin{equation}
V_{qq}(\vec{r}_{ij})=V_{CON}(\vec{r}_{ij})+V_{OGE}(\vec{r}_{ij})+V_{\chi}
(\vec{r}_{ij})+V_{S}(\vec{r}_{ij}) \,\,,
\label{int}
\end{equation}%
where the $i$ and $j$ indices are associated with $i$ and $j$ quarks
respectively, and ${\vec{r}}_{ij}$ stands for the interquark distance. 
$V_{\chi}$ denotes the pseudoscalar meson-exchange interaction
and $V_S$ stands for the scalar
meson-exchange potential described in Ref.~\cite{GFV07}.
Explicit expressions of all the
interacting potentials and a more detailed discussion 
of the model can be found in Refs.~\cite{Gar05,GFV07}.
In order to derive the local $B_1B_2\to B_3B_4$ potentials from the
basic $qq$ interaction defined above we use a Born-Oppenheimer
approximation. Explicitly, the potential is calculated as follows,

\begin{equation}
V_{B_1B_2 (L \, S \, T) \rightarrow B_3B_4 (L^{\prime}\, S^{\prime}\, T)} (R) =
\xi_{L \,S \, T}^{L^{\prime}\, S^{\prime}\, T} (R) \, - \, \xi_{L \,S \,
T}^{L^{\prime}\, S^{\prime}\, T} (\infty) \, ,  \label{Poten1}
\end{equation}

\noindent where

\begin{equation}
\xi_{L \, S \, T}^{L^{\prime}\, S^{\prime}\, T} (R) \, = \, {\frac{{\left
\langle \Psi_{B_3B_4 }^{L^{\prime}\, S^{\prime}\, T} ({\vec R}) \mid
\sum_{i<j=1}^{6} V_{qq}({\vec r}_{ij}) \mid \Psi_{B_1B_2 }^{L \, S \, T} ({\vec R%
}) \right \rangle} }{{\sqrt{\left \langle \Psi_{B_3B_4 }^{L^{\prime}\,
S^{\prime}\, T} ({\vec R}) \mid \Psi_{B_3B_4 }^{L^{\prime}\, S^{\prime}\, T} ({%
\vec R}) \right \rangle} \sqrt{\left \langle \Psi_{B_1B_2}^{L \, S \, T} ({\vec %
R}) \mid \Psi_{B_1B_2}^{L \, S \, T} ({\vec R}) \right \rangle}}}} \, .
\label{Poten2}
\end{equation}
In the last expression the quark coordinates are integrated out keeping $R$
fixed, the resulting interaction being a function of the $B_i-B_j$ 
relative distance. The wave function 
$\Psi_{B_iB_j}^{L \, S \, T}({\vec R})$ for the two-baryon system
is discussed in detail in Ref.~\cite{Val05}.

\subsection{Integral equations for the two-body systems}

If we consider the system of two baryons $B_1$ and $B_2$ 
with strangeness $-2$, $N \Xi $ and $Y_1Y_2$ ($Y_i=\Sigma, \Lambda$),
in a relative $S-$state interacting through a potential $V$ that contains a
tensor force, then there is a coupling to the $B_1B_2$ $D-$wave so that the
Lippmann-Schwinger equation for the spin singlet channels of Table~\ref{tA}
is of the form

\begin{eqnarray}
t_{\alpha\leftarrow \beta;ji}(p,p^{\prime \prime };W)
&=&V_{\alpha \leftarrow \beta;ji}(p,p^{\prime \prime })\nonumber \\
&+&\sum_{\gamma}\int_{0}^{\infty }{p^{\prime }}%
^{2}dp^{\prime }\,V_{\alpha \leftarrow \gamma;ji} (p,p^{\prime }) {\frac{2 \mu_\gamma}{k_\gamma^2 - p'^2 + i \epsilon}}%
t_{\gamma \leftarrow \beta;ji}(p^{\prime },p^{\prime \prime };W),  \label{eq1}
\end{eqnarray}
where $t$ is the two-body amplitude, $j$, $i$, and $W$ are the
total angular momentum, isospin and invariant energy of the system, and, for example,
for $i=0$ the subscripts $\alpha$, $\beta$, and $\gamma$ take the values
$\Lambda \Lambda$, $N \Xi$ and $\Sigma \Sigma$, and similarly for the
$i=1$ and $i=2$ channels shown in Table~\ref{tA}.

For the spin triplet channels of Table~\ref{tA} the 
Lippmann-Schwinger equation is of the form,

\begin{eqnarray}
t^{\ell \ell''}_{\alpha \leftarrow \beta;ji}(p,p^{\prime \prime };W)
&=&V^{\ell \ell''}_{\alpha \leftarrow \beta;ji}(p,p^{\prime \prime }) \nonumber \\
&+&\sum_{\gamma}\sum_{\ell'=0,2}\int_{0}^{\infty }{p^{\prime }}%
^{2}dp^{\prime }\,V^{\ell \ell'}_{\alpha \leftarrow \gamma;ji} (p,p^{\prime }) {\frac{2 \mu_\gamma}{k_\gamma^2 - p'^2 + i \epsilon}}%
t^{\ell' \ell''}_{\gamma \leftarrow \beta;ji}(p^{\prime },p^{\prime \prime };W),  \label{eq2}
\end{eqnarray}
with $\mu_\gamma$ the reduced mass of the system and the on-shell
momenta $k_\gamma$ are defined as
\begin{equation}
W=\sqrt{m_{1\gamma}^{2}+k_{\gamma}^{2}}+\sqrt{m_{2\gamma}^2+k_\gamma^2} \, ,
\label{forp10}
\end{equation}
where $m_1$ and $m_2$ are the masses of the particles of channel $\gamma$.

\subsection{Scattering cross sections}

We now turn to the available low-energy data on the $N\Xi $
scattering. There is only a small amount of data
corresponding to the total cross sections 
for $\Xi^{-}p\rightarrow \Xi^{-}p$, $\Xi^{-}p\rightarrow \Xi^{0}n$,
and $\Xi^{-}p\rightarrow\Lambda \Lambda$
reactions. 

In the case of processes of the type $N \Xi \rightarrow N \Xi$ the
amplitudes obtained from Eqs.~(\ref{eq1}) and~(\ref{eq2}) are related to
the cross section for a given isospin state through,
\begin{equation}
\sigma ^{i} = \pi^3\mu_{N\Xi}^2\left(3\mid t_{N\Xi \leftarrow N\Xi;1i}^{00}\mid^2
+\mid t_{N \Xi \leftarrow N \Xi;0i}^{00}\mid^2\right) \, .
\label{pup5}
\end{equation}
From the isospin cross sections the physical
channels are determined through,
\begin{eqnarray}
\sigma _{\Xi^{-}p\rightarrow \Xi^{-}p} &=& {\frac{1}{4}}\sigma ^{i=1}
+{\frac{1}{4}}\sigma^{i=0}+{\frac{1}{2}}\sqrt{\sigma^{i=0}\sigma^{i=1}} \,, \\
\sigma _{\Xi^{-}p\rightarrow \Xi^{0}n} &=& {\frac{1}{4}}\sigma ^{i=1}
+{\frac{1}{4}}\sigma^{i=0}-{\frac{1}{2}}\sqrt{\sigma^{i=0}\sigma^{i=1}} \,. \nonumber
\end{eqnarray}
In the case of the process $N\Xi \to\Lambda \Lambda$ it is necessary to
include also the transition with $\ell=2$ in the $\Lambda \Lambda$ channel.
Thus, in that case the cross section for isospin $i=0$ is
\begin{equation}
\sigma ^{0} = \pi^3\mu_{N\Xi}\mu_{\Lambda\Lambda}{k_{\Lambda \Lambda} \over k_{N\Xi}}
\left( \mid t_{\Lambda\Lambda \leftarrow N\Xi;00}^{00}\mid^2
+3 \mid t_{\Lambda\Lambda \leftarrow N\Xi;10}^{00}\mid^2
+3 \mid t_{\Lambda\Lambda \leftarrow N\Xi;10}^{02}\mid^2\right),
\label{pup6}
\end{equation}
and the cross section for the physical channel is
\begin{equation}
\sigma _{\Xi^{-}p\rightarrow \Lambda\Lambda} =\frac{1}{2}\sigma ^{0} \,. 
\label{pup7}
\end{equation}
Finally, for the elastic $\Lambda\Lambda \to \Lambda \Lambda$ process we have
\begin{equation}
\sigma _{\Lambda \Lambda \rightarrow \Lambda\Lambda} = 4 \pi^3 \mu_{\Lambda \Lambda}^2
\mid t_{\Lambda \Lambda \leftarrow \Lambda\Lambda;00}^{00}\mid^2 \,. 
\label{pup8}
\end{equation}

\section{Results}
\label{Res}

\begin{figure}[tbp]
\caption{$\Xi^- p$ elastic cross section, in mb, as a function of
the laboratory $\Xi$ momentum, in GeV/c. The two experimental data are taken
from Ref.~\cite{Tam01}, black circle, and Ref.~\cite{Yam01}, black square 
(both are in-medium experimental data). The solid
line indicates an upper limit for the cross section extracted in Ref.~\cite{Ahn06}
with a large uncertainty in the momentum.}
\mbox{\epsfxsize=140mm\epsffile{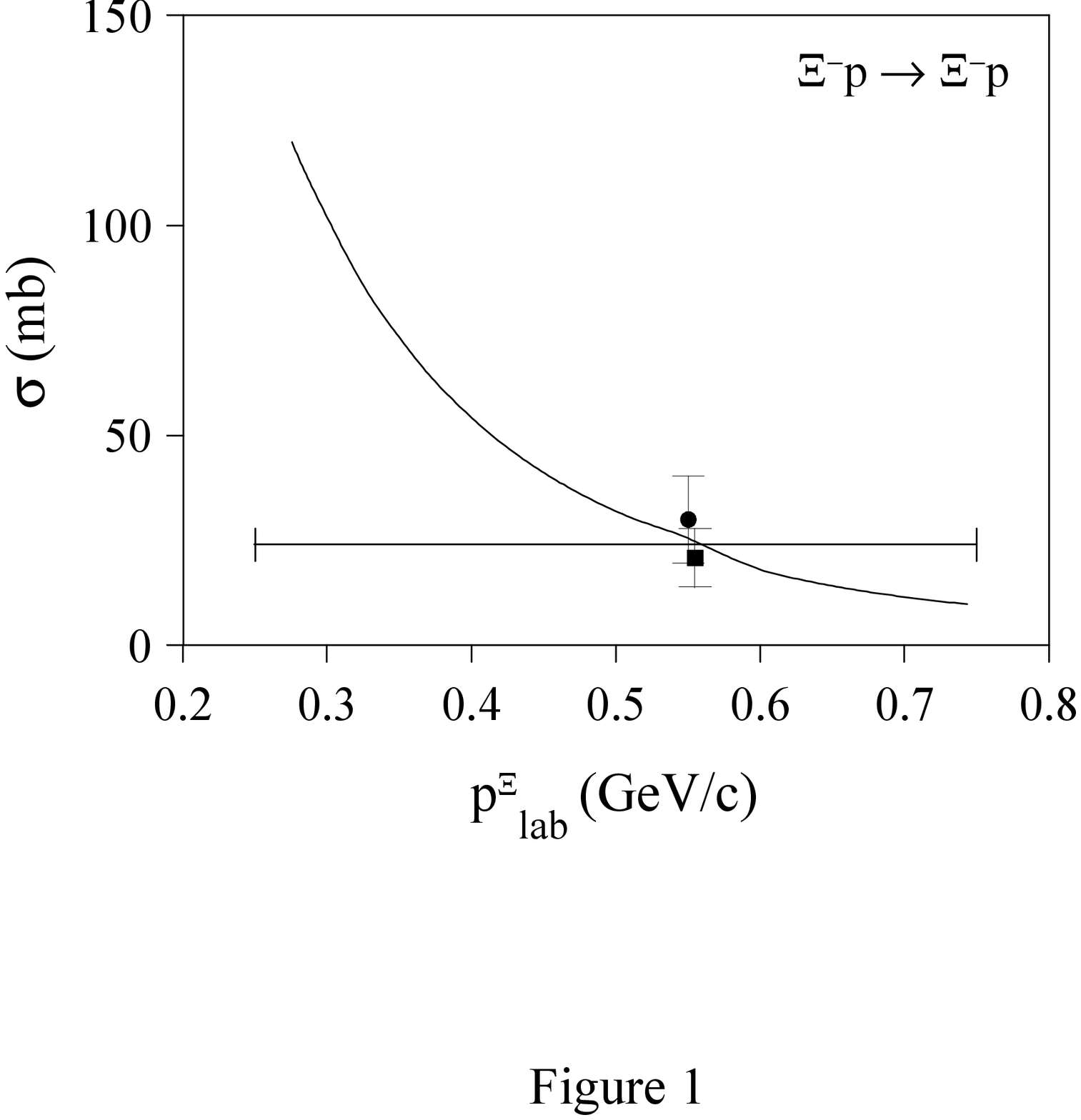}}
\label{fig1}
\end{figure}

\begin{figure}[tbp]
\caption{$\Xi^- p \to \Lambda\Lambda$ cross section, in mb, as a function of
the laboratory $\Xi$ momentum, in GeV/c. The experimental data is taken
from Ref.~\cite{Ahn06}.}
\mbox{\epsfxsize=140mm\epsffile{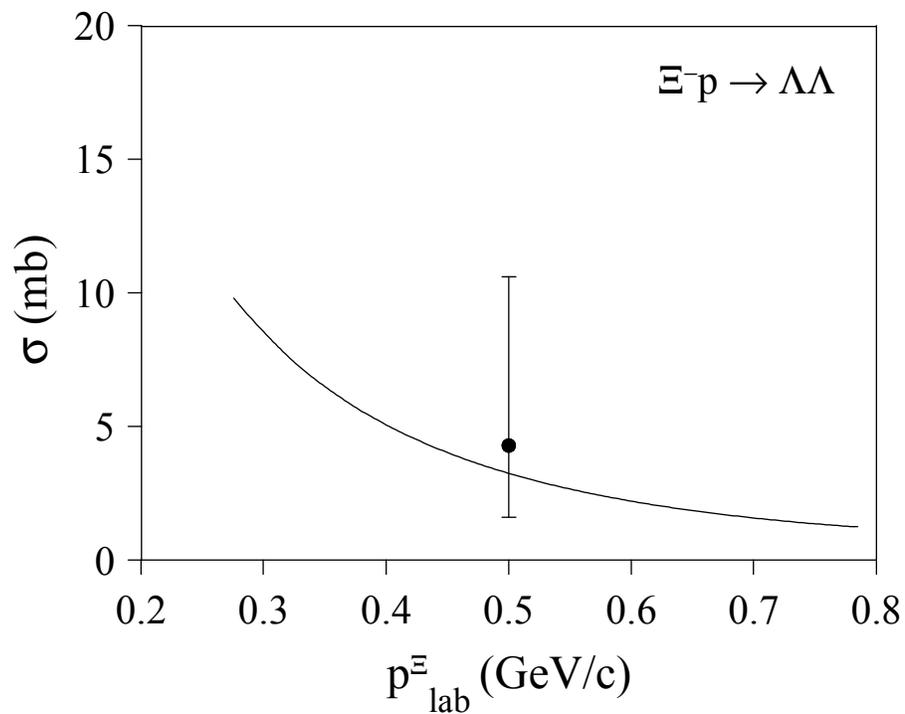}}
\label{fig2}
\end{figure}

\begin{figure}[tbp]
\caption{$\Xi^- p \to \Xi^0 n$ cross section, in mb, as a function of
the laboratory $\Xi$ momentum, in GeV/c. The experimental data
are taken from Ref.~\cite{Ahn06} as explained in the text.}
\mbox{\epsfxsize=140mm\epsffile{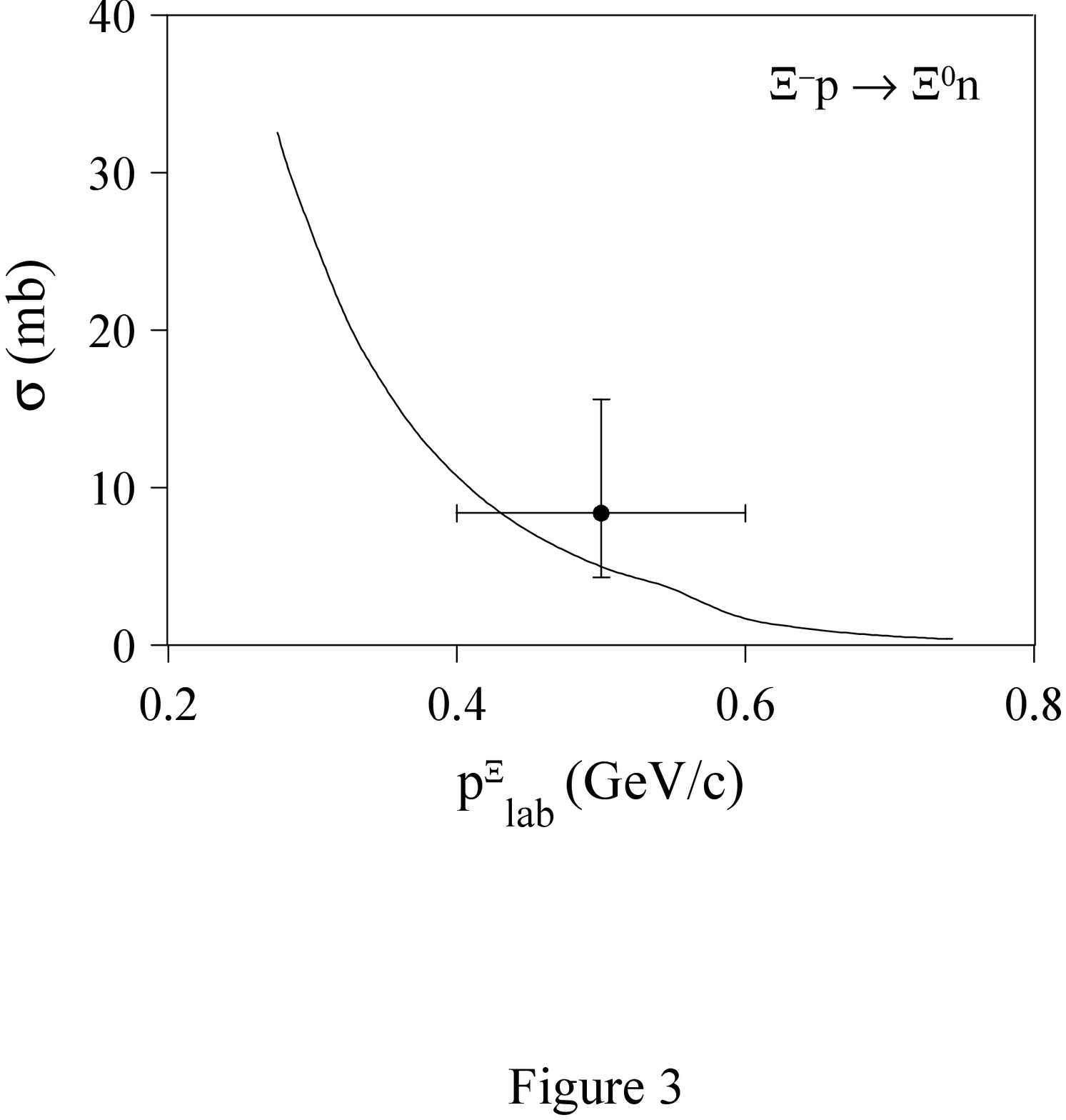}}
\label{fig3}
\end{figure}

\begin{figure}[tbp]
\caption{$\Lambda\Lambda$ cross section, in mb, as a function of
the laboratory $\Lambda$ momentum, in GeV/c.}
\mbox{\epsfxsize=140mm\epsffile{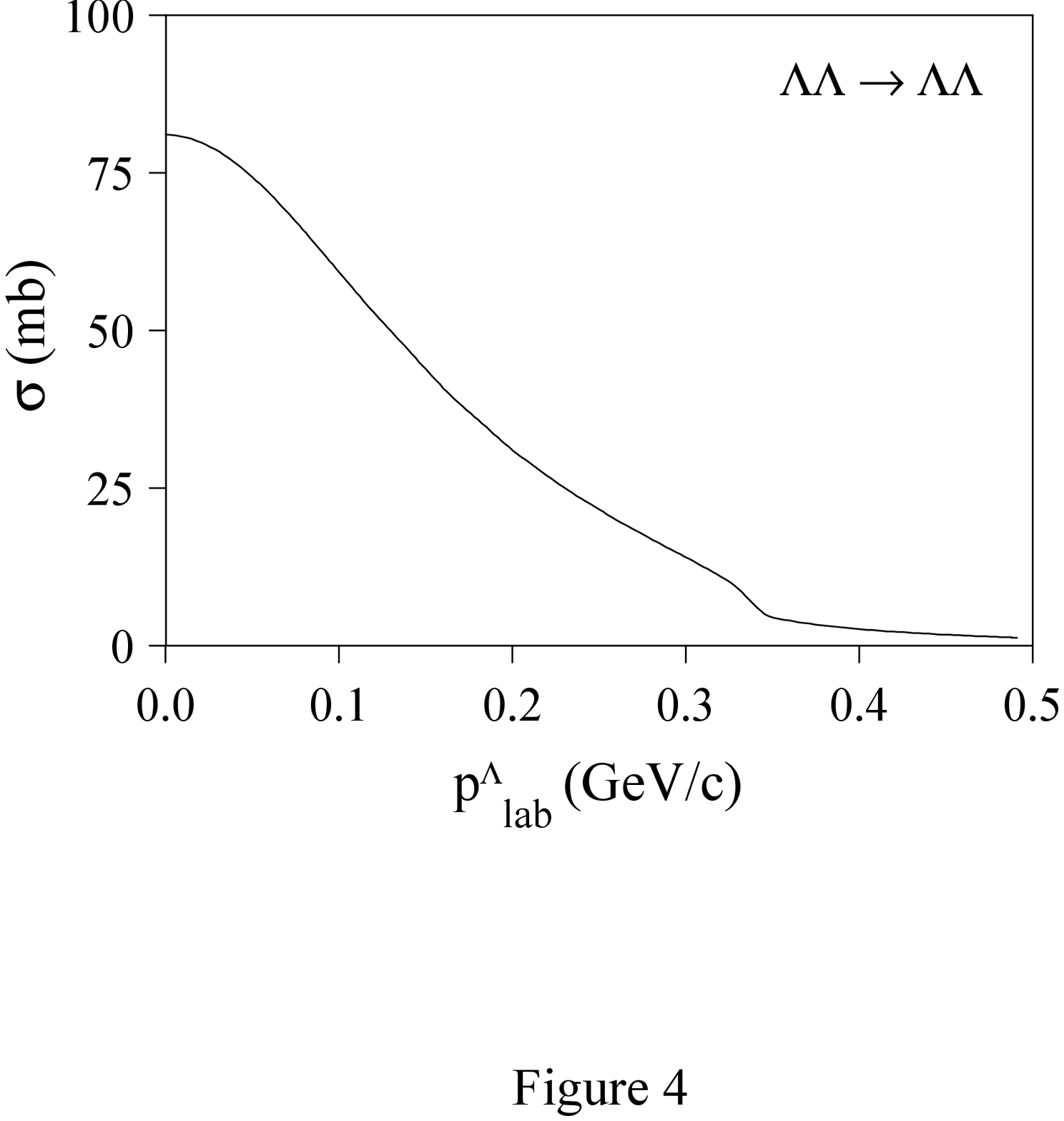}}
\label{fig4}
\end{figure}

Our results for the scattering cross sections are depicted by the solid lines in 
Figs.~\ref{fig1}, ~\ref{fig2}, ~\ref{fig3} and ~\ref{fig4}, compared to the
available experimental data. The scattering
lengths of the different spin-isospin channels are given in Table~\ref{tB}
compared to other theoretical models when available.

We show in Fig.~\ref{fig1} the $\Xi^- p$ elastic cross section compared
to the in-medium experimental $\Xi^- p$ cross section around $p^\Xi_{\rm lab}=$
550 MeV/c, where $\sigma_{\Xi^- p}=$ 30 $\pm$ 6.7 $^{+3.7}_{-3.6}$ mb~\cite{Tam01}.
Another analysis using the eikonal approximation gives
$\sigma_{\Xi^- p}=$ 20.9 $\pm$ 4.5 $^{+2.5}_{-2.4}$ mb~\cite{Yam01}. A more recent
experimental analysis~\cite{Ahn06} for the low energy $\Xi^- p$ elastic
and $\Xi^- p \to \Lambda \Lambda$ total cross sections in the range
0.2 GeV/c to 0.8 GeV/c shows that the former is less than 24 mb at 90\%
confidence level and the latter of the order of several mb, respectively.
In Fig.~\ref{fig2} we present the inelastic $\Xi^- p \to \Lambda\Lambda$
cross section. It has been recently estimated at a laboratory momentum
of $p^\Xi_{\rm lab}=$500 MeV/$c$, see Ref.~\cite{Ahn06}, assuming a quasifree 
scattering process for the reaction $^{12}{\rm C}(\Xi^-,\Lambda\Lambda)X$ 
obtaining a total cross section
$\sigma(\Xi^- p \to \Lambda \Lambda)=4.3^{+6.3}_{-2.7}$ mb. 
The upper limit of the cross section was derived as 12 mb at 90\% confidence level.
Fig.~\ref{fig3} shows the total inelastic cross section $\Xi^- p \to \Xi^0 n$.
Combining the results of Refs.~\cite{Ahn06} and ~\cite{Ahn98}, one obtains
$\sigma(\Xi^- p \to \Xi^0 n) \sim 10$ mb. A recent measurement of a 
quasifree $p(K^-,K^+)\Xi^-$ reaction in emulsion plates yielded
$12.7^{+3.5}_{-3.1}$ mb for the total inelastic cross section in
the momentum range $0.4-0.6$ GeV/$c$~\cite{Aok98}, consistent with
the results of Ref.~\cite{Ahn98}. The experimental results for
$\Xi^- p$ inelastic scattering of Refs.~\cite{Ahn98,Aok98} involve
both $\Xi^- p \to \Lambda \Lambda$ and $\Xi^- p \to \Xi^ 0 n$~\cite{Ahn06},
what combined with the value for $\Xi^- p \to \Lambda\Lambda$ of Ref.~\cite{Ahn06}
allows to obtain an estimate of the inelastic cross section $\Xi^- p \to \Xi^0 n$.
Finally in Fig.~\ref{fig4} we present our prediction for the $\Lambda \Lambda$
scattering cross section. In this last case there are no experimental data.

As can be seen our results agree with the
experimental data for the elastic and inelastic $\Xi N$
cross sections. The small bumps in the cross sections
correspond to the opening of inelastic channels.
As explained above, the interacting model is taken for grant from
Ref.~\cite{GFV07}, where three body systems
with strangeness $-1$ were studied, involving therefore
two-body subsystems with strangeness $0$ and $-1$. The 
agreement with the experimental data gives
support to the dynamical model used and make
our predictions valuable for forthcoming experiments. 
We would like to emphasize the agreement of our
results with the $\Xi^- p \to \Lambda \Lambda$ conversion cross section.
This reaction is of particular importance in assessing the stability
of $\Xi^-$ quasi-particle states in nuclei. 
Our results are close to the estimations of the Nijmegen-D model~\cite{NijD}.
Ref.~\cite{Fuj07} predicts $\sigma (\Xi^- p \to \Xi^0 n)\sim$
15 mb at $p^\Xi_{\rm lab}=$0.5 GeV/$c$. Ref.~\cite{Ahn98}
reported 14 mb for the inelastic scattering involving
both $\Xi^- p \to \Lambda \Lambda$ and $\Xi^- p \to \Xi^0 n$.
We found a smaller 
value of approximately 6 mb for both inelastic channels,
in close agreement to experiment.
In the case of the elastic $\Lambda\Lambda$ cross section
there are no experimental data. Our predictions are rather 
similar to Refs.~\cite{Pol07,Fuj07}.
We definitely need more experimental data with high statistics. 
This will help us in discriminating among the different dynamical
models. A theoretical evaluation of the in-medium cross sections 
would also be necessary.

The scattering lengths for the different spin-isospin
channels are given in Table~\ref{tB}. These parameters are complex 
for the $N\Xi$ $(i,j)=(0,0)$ since the inelastic $\Lambda \Lambda$
channel is open, for the $\Lambda \Sigma$ isospin 1 channel due
to the opening of the $N\Xi$ channel, and for the 
$\Sigma\Sigma$ $i=$0 and 1 since the $\Lambda\Lambda$
and $N\Xi$ channels are open in the first case, and the $N\Xi$ and $\Lambda\Sigma$
are open in the second (see Table~\ref{tA}).  
There is no direct comparison between our results and those of
Ref.~\cite{Fuj07}. Although both are quark-model based results,
Ref.~\cite{Fuj07} used an old-fashioned quark-model interacting potential.
For example, they use a huge
strong coupling constant, $\alpha_S=1.9759$, and they consider the contribution of
vector mesons what could give rise to double counting problems~\cite{Yaz90}. 
As explicitly written in Ref.~\cite{Fuj07}, the parameters
are effective in their approach and has very little to do with QCD.
As mentioned above, since the observation
of the Nagara event~\cite{Nag08} it is generally accepted
that the $\Lambda\Lambda$ interaction is only moderately attractive.
Our result for the $^1S_0$ $\Lambda\Lambda$ scattering length is
compatible with such event. A rough estimate of $B_{\Lambda\Lambda}(
B_{\Lambda\Lambda}\approx (\hbar c)^2 / 2 \mu_{\Lambda\Lambda} {a^{\Lambda\Lambda}_{^1S_0}}^2$)
drives a value of 5.41 MeV, below the upper limit extracted from the Nagara
event, $B_{\Lambda\Lambda}=7.25\pm0.19$ MeV. Moreover, although from
the $\Lambda\Lambda$ scattering length alone one cannot draw any conclusion
on the magnitude of the two-$\Lambda$ separation energy, recent 
estimates~\cite{Rij06b} have reproduced the two-$\Lambda$ separation
energy, defined as $\Delta B_{\Lambda\Lambda}=B_{\Lambda\Lambda}(^6_{\Lambda\Lambda}{\rm He})
-2B_{\Lambda}(^5_\Lambda {\rm He})$, with scattering lengths of $-1.32$ fm.
The only reliable way to determine the two-$\Lambda$ separation energy
in our model would be a concrete calculation of doubly-strange hypernuclei.
This has not been done so far, and it is not clear that such calculation
might help to further constrain the potential model.

\section{Summary and outlook}
\label{Sum}

In this letter we have presented the first results for the doubly strange
$N \Xi $ and $Y_1Y_2$ interactions ($Y_i=\Sigma, \Lambda$) obtained with
a constituent quark model approach designed to study the non-strange
hadron phenomenology. The interaction incorporates long-ranged
meson-exchange contributions and the short-range dynamics generated
by the one-gluon exchange and quark antisymmetry contributions.

We showed that the CCQM predictions are consistent with the recently
obtained doubly strange elastic and inelastic 
scattering cross sections. In particular, our results are compatible with the
$\Xi^- p \to \Lambda \Lambda$ conversion cross section,
important in assessing the stability
of $\Xi^-$ quasi-particle states in nuclei.
Furthermore a moderately attractive $\Lambda\Lambda$ interaction 
arises. The presently available scattering data
are, however, not sufficient to draw definitive conclusions about
the model, but forthcoming experimental data of the observables 
reported will help in testing different theoretical model predictions.

It is expected that in the coming years better-quality data on the
fundamental $N \Xi$ and $YY$ interactions as well as much information
about the physics of hypernuclei will become available at the new
facilities J-PARC (Tokai, Japan) and FAIR (Darmstadt, Germany).
The CCQM developed here can then be used to analyze these 
upcoming data in a model-independent way. 

\acknowledgments
The authors greatly acknowledge C.~J.~Yoon for useful information
about the results of Ref.~\cite{Yoo07}.
This work has been partially funded by Ministerio de Ciencia y Tecnolog\'{\i}a
under Contract No. FPA2007-65748 and by EU FEDER, by Junta de Castilla y Le\'{o}n
under Contract No. GR12,
by the Spanish Consolider-Ingenio 2010 Program CPAN (CSD2007-00042),
by HadronPhysics2, a FP7-Integrating Activities and Infrastructure
Program of the European Commission, under Grant 227431, and
by COFAA-IPN (M\'{e}xico).

\newpage

\begin{table}[tbp]
\caption{Interacting channels on the spin($j$)-isospin($i$) basis
for the $S-D$ partial waves.}
\label{tA}
\begin{tabular}{|c|ccc|}
\hline
& $i=0$  &  $i=1$ & $i=2$ \\
\hline\hline
$j=0$ & $\Lambda\Lambda - N\Xi - \Sigma\Sigma$ & $N\Xi - \Lambda\Sigma$ & $\Sigma\Sigma$ \\
$j=1$ & $N\Xi$ & $N\Xi - \Lambda\Sigma - \Sigma\Sigma$ & $-$ \\ \hline
\end{tabular}
\end{table}

\begin{table}[tbp]
\caption{Two-body singlet and triplet scattering lengths, in fm, for different
models as compared to our results. The results between squared brackets indicate the
lower and upper limit for different parametrizations used in that reference.}
\label{tB}
\begin{tabular}{|c|ccccc|}
\hline
Model & Ref.~\cite{Sto99} & Ref.~\cite{Rij06} & Ref.~\cite{Pol07} & Ref.~\cite{Fuj07} & Ours \\
\hline\hline
$a_{^1S_0}^{\Lambda\Lambda}$  & $[-0.27,-0.35]$ & $[-1.555,-3.804]$ & $[-1.52,-1.67]$ & $-0.821$ & $-2.54$ \\
$a_{^1S_0}^{\Xi^0 p}$         & $[0.40,0.46]$   & $[0.144,0.491]$  & $[0.13,0.21]$  & $0.324 $ & $-3.32$ \\
$a_{^3S_1}^{\Xi^0 p}$         & $[-0.030,0.050]$  & $-$   & $[0.0,0.03]$   & $-0.207$ & $18.69$ \\
$a_{^1S_0}^{\Sigma^+\Sigma^+}$& $[6.98,10.32]$  & $-$           & $[-6.23,-9.27]$& $-85.3$      & $0.523$ \\ \hline
$a_{^1S_0}^{\Xi N (I=0)}$     &$   -       $   & $     -        $  & $   -       $  & $  -  $ & $-1.20\, + \, i \, 0.75$ \\
$a_{^3S_1}^{\Xi N (I=0)}$         &$   -        $  & $[-1.672,122.5]$   & $   -      $   & $  -   $ & $0.28$ \\
$a_{^1S_0}^{\Lambda\Sigma}$   &$   -       $   & $     -        $  & $   -       $  & $  - $ & $0.908\, + \, i \, 1.319$ \\
$a_{^3S_1}^{\Lambda\Sigma}$   &$   -   $   & $     -   $  & $   -       $  & $  - $ & $ - $ $3.116\, + \, i \, 0.393$ \\
$a_{^3S_1}^{\Sigma\Sigma (I=1)}$   &$   -   $   & $     -   $  & $   -       $  & $  - $ & $ - $ $1.347\, + \, i \, 1.801$ \\
$a_{^1S_0}^{\Sigma\Sigma (I=0)}$   &$   -   $   & $     -   $  & $   -       $  & $  - $ & $ - $ $0.039\, + \, i \, 0.517$ \\ \hline
\end{tabular}
\end{table}

\newpage

\newpage

\begin{thebibliography}{}

\bibitem{Nag08} H.~Takahashi {\it et al.},
			Phys. Rev. Lett. {\bf 87}, 212502 (2001).

\bibitem{Nak09} K.~Nakazawa {\it et al.}, Proceedings of the 10th
International Conference on Hypernuclear and Strange Particle
Physics (HypX), Tokai (2009).

\bibitem{Ich01} A.~Ichikawa, Ph.D. Thesis, Kyoto University (2001).

\bibitem{Hiy10} E.~Hiyama, M.~Kamimura, Y.~Yamamoto, and T.~Motoba,
			Phys. Rev. Lett. {\bf 104}, 212502 (2010).

\bibitem{Ahn06} J.~K.~Ahn {\it et al.},
			Phys. Lett. B {\bf 633}, 214 (2006). 

\bibitem{Yoo07} C.~J.~Yoon {\it et al.},
			Phys. Rev. C {\bf 75}, 022201(R) (2007).

\bibitem{Nag00} T.~Nagae {\it et al.}, J-PARC proposal E05,
$http://j-parc.jp/NuclPart/pac_0606/pdf/p05-Nagae.pdf$.

\bibitem{Har10} T.~Harada, Y.~Hirabayashi, and A.~Umeya,
			Phys. Lett. B {\bf 690}, 363 (2010). 

\bibitem{Val05} A.~Valcarce, H.~Garcilazo, F.~Fern\'andez, and P.~Gonz\'alez,
			Rep. Prog. Phys. {\bf 68}, 965 (2005).

\bibitem{Gar05} A.~Valcarce, H.~Garcilazo, and J.~Vijande,
			Phys. Rev. C {\bf 72}, 025206 (2005).
                J.~Vijande, F.~Fern\'andez, and A.~Valcarce, 
			J. Phys. G {\bf 31}, 481 (2005).

\bibitem{GFV07} H.~Garcilazo, T.~Fern\'andez-Caram\'es, and A.~Valcarce, 
			Phys. Rev. C {\bf 75}, 034002 (2007).

\bibitem{Tam01} T.~Tamagawa {\it et al.},
			Nucl. Phys. A {\bf 691}, 234c (2001).
			
\bibitem{Yam01} Y.~Yamamoto, T.~Tamagawa, T.~Fukuda, and T.~Motoba,
			Prog. Theor. Phys. {\bf 106} 363 (2001). 


\bibitem{Ahn98} J.~K.~Ahn {\it et al.},
			Nucl. Phys. A {\bf 625}, 231 (1997). 

\bibitem{Aok98} S.~Aoki {\it et al.},
			Nucl. Phys. A {\bf 644}, 365 (1998). 

\bibitem{Fuj07} Y.~Fujiwara, Y.~Suzuki, and C.~Nakamoto,
			Prog. Part. Nucl. Phys. {\bf 58}, 439 (2007). 

\bibitem{Pol07} H.~Polinder, J.~Haidenbauer, and U.~-G.~Meissner,
			Phys. Lett. B {\bf 653}, 29 (2007). 

\bibitem{Sto99} V.~G.~J.~Stoks and Th.~A.~Rijken,
                        Phys. Rev. C {\bf 59}, 3009 (1999).

\bibitem{Rij06} Th.~A.~Rijken and Y.~Yamamoto,
			nucl-th/0608074.

\bibitem{NijD} C.~Nakamoto, Y.~Fujiwara, and Y.~Suzuki,
			Nucl. Phys. A {\bf 639}, 51c (1998). 

\bibitem{Yaz90} K.~Yazaki,
			Prog. Part. Nucl. Phys. {\bf 24}, 353 (1990). 

\bibitem{Rij06b} Th.~A.~Rijken and Y.~Yamamoto,
                        Phys. Rev. C {\bf 73}, 044008 (2006).


\end{thebibliography}
\end{document}